\RequirePackage{lineno}
\documentclass[twocolumn,aps,prc,showpacs,superscriptaddress,floatfix]{revtex4}
\usepackage{epsfig}
\usepackage{color}
\usepackage{lineno}
\newcommand {\snn}	{\sqrt{s_{_{\rm NN}}}}
\newcommand {\gevc}	{GeV/$c$}
\newcommand {\fmc}	{fm/$c$}
\newcommand {\AuAu}	{Au+Au}
\newcommand {\dAu}	{$d$+Au}
\newcommand {\pPb}	{$p$+Pb}
\newcommand {\bAu}	{$b=6.6$-8.1~fm}
\newcommand {\bzero}	{$b=0$~fm}
\newcommand {\pbar}	{\bar{p}}
\newcommand {\pt}	{p_{\perp}}
\newcommand {\rt}	{r_{\perp}}
\newcommand {\dphi}	{\Delta\phi}
\newcommand {\ncq}	{n_{\rm q}}
\newcommand {\phirndm}	{$\phi$-randomized}
\newcommand {\tmax}	{t_{\rm max}}
\newcommand {\eini}	{\epsilon_2}
\newcommand {\ehad}	{\epsilon_2^{\rm had}}
\newcommand {\vini}	{v_2^{0.6{\rm fm}/c}}
\newcommand {\vfin}	{v_2^{30{\rm fm}/c}}
\newcommand {\dv}	{\Delta v_2}
\newcommand {\mean}[1]	{\langle #1\rangle}

\begin{document}
%%%%%%%%%%%%%%%%%%%%%%%%%%%%%%%%%%%%%%%%%%%%%%%%%%%%%%%%%%%%%%%
%\linespread{1.6}
\title{Origin of the mass splitting of azimuthal anisotropies in a multi-phase transport model}
\author{Hanlin Li}
\address{College of Science, Wuhan University of Science and Technology, Wuhan, Hubei 430065, China}
\address{Department of Physics and Astronomy, Purdue University, West Lafayette, Indiana 47907, USA}
\author{Liang He}
\address{Department of Physics and Astronomy, Purdue University, West Lafayette, Indiana 47907, USA}
\author{Zi-Wei Lin}
\address{Department of Physics, East Carolina University, Greenville, North Carolina 27858, USA}
\address{Key Laboratory of Quarks and Lepton Physics (MOE) and Institute of Particle Physics, Central China Normal University, Wuhan, Hubei 430079, China}
\author{Denes Molnar}
\address{Department of Physics and Astronomy, Purdue University, West Lafayette, Indiana 47907, USA}
\author{Fuqiang Wang}
\email{fqwang@purdue.edu}
\address{Department of Physics and Astronomy, Purdue University, West Lafayette, Indiana 47907, USA}
\address{School of Science, Huzhou University, Huzhou, Zhejiang 313000, China}
\author{Wei Xie}
\address{Department of Physics and Astronomy, Purdue University, West Lafayette, Indiana 47907, USA}
\date{\today}

\begin{abstract}
Both hydrodynamics-based models and a multi-phase transport  (AMPT) model can reproduce 
the mass splitting of azimuthal anisotropy ($v_n$) at low transverse momentum ($\pt$) as observed in heavy ion collisions. In the AMPT model, however, $v_n$ is mainly generated by the parton escape mechanism, not by the hydrodynamic flow.
In this study we provide detailed results on the mass splitting of $v_n$ in this transport model, including $v_2$ and $v_3$ of various hadron species in \dAu\ and Au+Au collisions at the Relativistic Heavy Ion Collider and \pPb\ collisions at the Large Hadron Collider. 
We show that the mass splitting of hadron $v_2$ and $v_3$ in AMPT first arises from the kinematics in the quark coalescence hadronization process, and then, more dominantly, comes from hadronic rescatterings, even though the contribution from the latter to the overall charged hadron $v_n$ is small.  
We further show that there is no qualitative difference between heavy ion collisions and small-system collisions or between elliptic ($v_2$) and triangular ($v_3$) anisotropies. Our studies thus demonstrate that the mass splitting of $v_2$ and $v_3$ at low-$\pt$ is not a unique signature of hydrodynamic collective  flow but can be the interplay of several physics effects.
\end{abstract}
%\begin{keyword}
%quark-gluon plasma \sep anisotropic flow \sep transport model \sep hydrodynamics
%\PACS 25.75.-q \sep 25.75.Ld
%\end{keyword}
\maketitle
%\linespread{1.6}

%%%%%%%%%%%%%%%%%%%%%%%%%%%%%%%%%%%%%%%%%%%%%%%%%%%%%%%%%%%%%%%
\section{Introduction}

The quark-gluon plasma has been created in relativistic heavy ion collisions, and 
extensive efforts are going on to study quantum chromodynamics at the extreme conditions of high temperature and energy density~\cite{Arsene:2004fa,Back:2004je,Adams:2005dq,Adcox:2004mh,Muller:2012zq}. 
Of particular interests are non-central heavy ion collisions, where the overlap volume of the colliding nuclei is anisotropic in the transverse plane perpendicular to the beam direction. 
One interesting finding is that the collision system is explosive, consistent with the buildup and expansion of the hydrodynamic pressure~\cite{Heinz:2013th,Gale:2013da,Abelev:2008ab}.
The pressure gradient and/or particle interactions would generate an anisotropic expansion, which converts the anisotropic geometry into the final-state elliptic flow~\cite{Ollitrault:1992bk}.
In addition, due to fluctuations in the initial-state collision geometry, there is an elliptic harmonic anisotropy in the configuration space $(\epsilon_2$) even in central heavy ion or proton-nucleus collisions~\cite{Andrade:2006yh}. Furthermore, fluctuations lead to finite configuration space harmonics of all orders~\cite{Alver:2010gr}, which will result in final-state momentum anisotropies of all orders ($v_n$), where $n$ is a positive integer.

The mass splitting of hadron $v_2$ at low transverse momentum ($\pt$) is also observed in the experimental data. It is often considered as a hallmark of the hydrodynamic description of relativistic heavy ion collisions~\cite{Heinz:2013th}, where a common but anisotropic transverse velocity field coupled  with the Cooper-Frye hadronization mechanism~\cite{Cooper:1974mv} leads to the mass splitting. Furthermore, results from hybrid models, where hydrodynamics is followed by a hadron cascade, have shown that the $v_2$ mass splitting is small just after hadronization and is then strongly enhanced by hadronic  scatterings~\cite{Hirano:2007ei,Song:2010aq,Romatschke:2015dha}. 

Large $v_n$ values have been observed in large-system heavy ion collisions, and both hydrodynamics-based models~\cite{Heinz:2013th,Gale:2013da,Abelev:2008ab} and a multi-phase transport (AMPT) model~\cite{Lin:2001zk,Lin:2004en,Lin:2014tya} can reproduce these results. 
Later particle correlation data in small systems, including \dAu~\cite{Adare:2014keg,Adamczyk:2015xjc} collisions at the Relativistic Heavy Ion Collider (RHIC) and high multiplicity $p$+$p$~\cite{Khachatryan:2010gv} or $p$+Pb~\cite{CMS:2012qk,Abelev:2012ola,Aad:2012gla} collisions at the Large Hadron Collider (LHC), 
hint at similar $v_n$ (and mass splitting).
Again, both hydrodynamics~\cite{Bozek:2010pb,Bozek:2012gr} and a multi-phase transport~\cite{Bzdak:2014dia} can reasonably describe the experimental data. This seems puzzling, because naively one would expect the small system to be far from equilibrium and thus not suitable for a hydrodynamical description. 

A recent study by some of us~\cite{He:2015hfa,Lin:2015ucn} using AMPT has shown that the azimuthal anisotropy is mainly generated by the anisotropic parton escape and that hydrodynamics may play only a minor role. This escape mechanism would naturally explain the similar azimuthal anisotropies in heavy ion and small system collisions. Since mass splitting of hadron $v_2$ is also present in the AMPT results, it suggests that the hydrodynamic collective flow may not be the only mechanism that can generate the mass splitting of hadron $v_n$ in collisions with high energy densities.

In an earlier study~\cite{Li:2016flp} we used AMPT simulations of Au+Au and \dAu\ collisions at the top RHIC energy to investigate the mass splitting of $v_2$ of pions, kaons, and protons. 
We found that the mass splitting of $v_2$ in AMPT is partly due to the kinematics in the quark coalescence process but mainly due to hadronic rescatterings~\cite{Li:2016flp}. 
In this paper we expand that study to more hadron species including $\rho$, $K^*$, $\phi$, $\Delta$ and strange hadrons such as $\Lambda$ and $\Xi$. 
We also investigate the massing splitting of the triangular flow $v_3$ and include AMPT results of \pPb\ collisions at the LHC energy of 5~TeV.
In addition, we provide details of our analysis, such as the effect of the finite opening angles among coalescing partons, the difference between primordial hadrons and hadrons from resonance decays, and the connection between the $v_2$ mass splitting and the initial hadron spatial eccentricity.

\section{Model and Analysis}

We employ the same version of the string melting AMPT model
(v2.26t5, available online at~\cite{AMPTcode}) as in earlier studies ~\cite{He:2015hfa,Lin:2015ucn,Li:2016flp}. 
It consists of a fluctuating initial condition, parton elastic scatterings, quark coalescence for hadronization, and hadronic interactions. 
The initial energy and particle productions are being described by the HIJING model. 
However, the string melting AMPT model converts these initial hadrons to their valence quarks and antiquarks, based on the assumption that the high energy density in the overlap region of high energy heavy ion collisions requires us to use parton degrees of freedom to describe the dense matter~\cite{Lin:2001zk}. Two-body elastic parton scatterings are treated with Zhang's Parton Cascade (ZPC)~\cite{Zhang:1997ej}, 
where we take the strong coupling constant $\alpha_s=0.33$ and a total parton scattering cross section $\sigma=3$~mb for all AMPT calculations in this study. 
After partons stop interacting, a simple quark coalescence model is applied to describe the hadronization process that converts partons into hadrons ~\cite{Lin:2004en}. Subsequent interactions of these formed hadrons are modeled by a hadron cascade~\cite{Lin:2004en}. 

Two of the above components, the hadronization process and hadron cascade, are especially relevant for this study. 
Hadronization in the string melting version of AMPT is modeled with a simple quark coalescence, where two nearest partons in space (one quark and one antiquark) are combined into a meson and three nearest quarks (or antiquarks) are combined into a baryon (or antibaryon). In addition, when the flavor composition of the coalescing quark and antiquark allows the formation of either a pseudo-scalar or a vector meson, the meson species whose mass is closer to the invariant mass of the coalescing parton pair will be formed. The same criterion is also applied to the formation of an octet or a decuplet baryon with the same flavor composition. Thus in these situations the hadron species that has a larger mass will be typically formed when the coalescing partons have a larger invariant mass.

The hadron cascade in the AMPT model includes explicit particles such as $\pi$, $\rho$, $\omega$, $\eta$, $K$, $K^*$, $\phi$ mesons, $N$, $\Delta$, $N^*(1440)$, $N^*(1535)$, $\Lambda$, $\Sigma$, $\Xi$, $\Omega$, and deuteron and the corresponding anti-particles~\cite{Oh:2009gx}. 
Hadronic interactions include meson-meson, meson-baryon, and baryon-baryon elastic and inelastic scatterings. For example, meson-baryon scatterings includes pion-nucleon, $\rho$-nucleon, and kaon-nucleon elastic and inelastic processes, among many reaction channels. More details can be found in the main AMPT paper \cite{Lin:2004en}. We terminate the hadronic interactions at a cutoff time ($\tmax$), when the observables of interest are stable; a default cutoff time of $\tmax=30$~\fmc\ is used here. 

In this study we simulate three collision systems: \AuAu\ collisions at RHIC with \bAu\ (corresponding to approximately 20\%-30\% centrality~\cite{Abelev:2008ab}) at the nucleon-nucleon center-of-mass energy $\snn=200$~GeV, \dAu\ collisions at RHIC with \bzero\ at $\snn=200$~GeV, and \pPb\ collisions at LHC with \bzero\ at $\snn=5$~TeV.
Note that the string melting version of AMPT can reasonably reproduce the particle yields, $\pt$ spectra, and  $v_2$ of low-$\pt$ pions and kaons in central and mid-central \AuAu\ collisions at 200A GeV and Pb+Pb collisions at 2760A GeV~\cite{Lin:2014tya}. 

The initial geometric anisotropy of the transverse overlap region of a heavy-ion collision is often described by the eccentricity of the $n$th harmonic order~\cite{Alver:2010gr}:
\begin{equation}
  \epsilon_n=\left.\sqrt{\mean{\rt^2\cos n\phi_r}^2+\mean{\rt^2\sin n\phi_r}^2}\right/\mean{\rt^2}\,.
\end{equation}
Here $\rt$ and $\phi_r$ are the polar coordinate of each initial parton (after its formation time) in the transverse plane, and $\mean{...}$ denotes the per-event average. 
We follow the same method as in our earlier studies~\cite{He:2015hfa,Li:2016flp}
to calculate azimuthal anisotropies.
In particular, we compute the $n^{\rm th}$ harmonic plane 
(short-axis direction of the corresponding harmonic component) of each event 
from its initial configuration of all partons~\cite{Ollitrault:1993ba} according to
\begin{equation}
\psi_n^{(r)}=\frac{1}{n}\left[{\rm atan2}(\mean{\rt^{2}\sin n\phi_r},\mean{\rt^{2}\cos n\phi_r})+\pi\right]\,.
\end{equation}
The momentum anisotropies are then characterized by Fourier coefficients~\cite{Voloshin:1994mz}
\begin{equation}
v_n^{\rm obs}=\mean{\cos n(\phi -\psi_n^{(r)})}\,,
\end{equation}
where $\phi$ is the azimuthal angle of the parton or hadron momentum.
Note that all results shown in this paper are for particles (partons or hadrons) within the pseudo-rapidity window of $|\eta|<1$.

%%%%%%%%%%%%%%%%%%%%%%%%%%%%%%%%%%%%%%%%%%%%%%%%%%%%%%%%%%%%%%%
\section{Partonic anisotropy}

\begin{figure*}[hbt]
  \begin{center}
    \includegraphics[width=\textwidth]{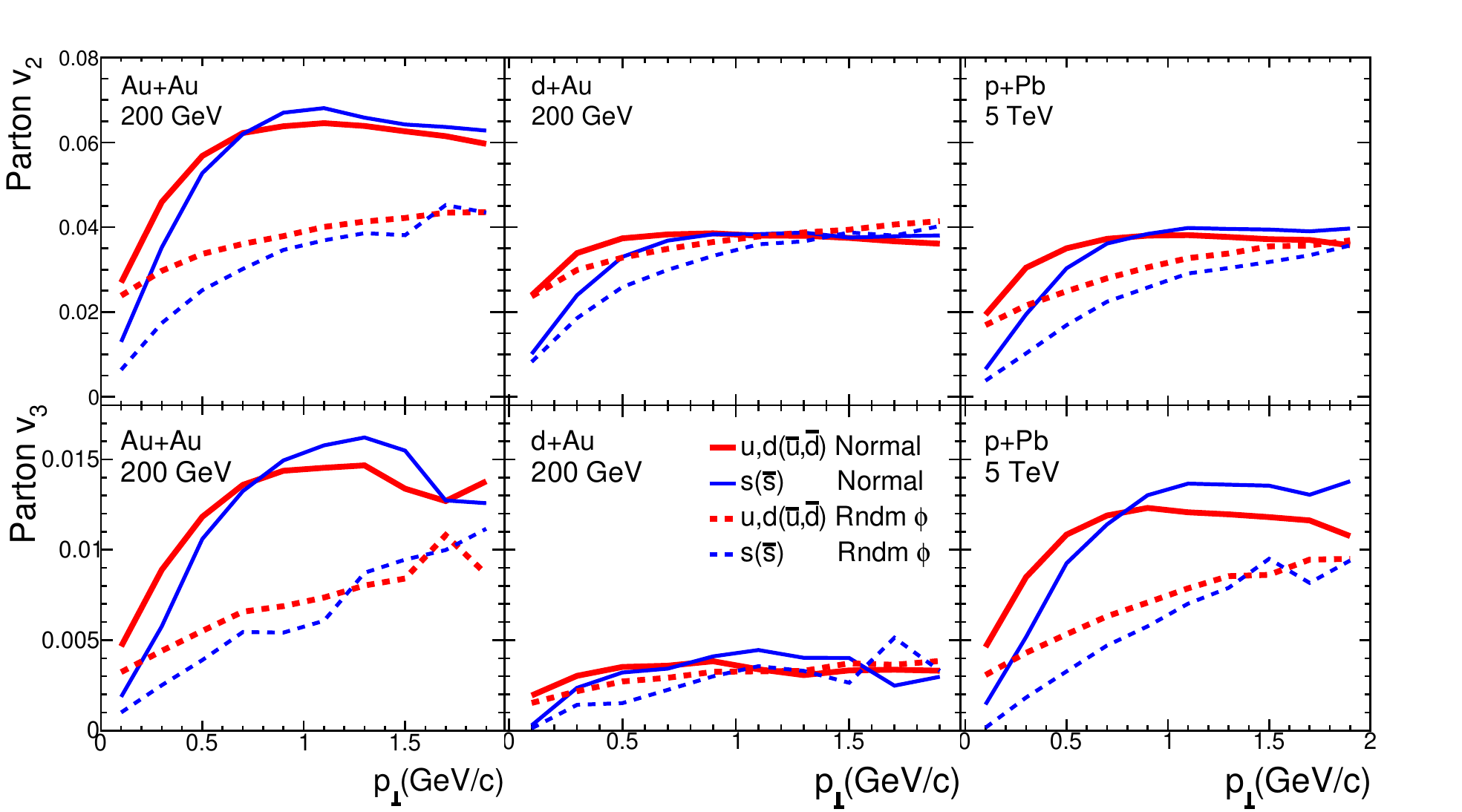}
    \caption{(Color online)  {\em Parton $v_n$.} Parton $v_2$ (upper panels) and $v_3$ (lower panels) as a function of $\pt$ for light ($u$ and $d$) and strange ($s$) (anti-)quarks in the final state before hadronization from AMPT with string melting. Three systems are shown: \bAu\ \AuAu\ collisions at $\snn=200$~GeV (left column), \bzero\ \dAu\ collisions at 200~GeV (middle column), and \bzero\ \pPb\ at 5~TeV (right column). Both normal (solid curves) and \phirndm\ (dashed curves) AMPT results are shown, with thick curves for light quarks and thin curves for strange quarks.}
    \label{fig:uds}
  \end{center}
\end{figure*}

Currently the string melting version of the AMPT model ~\cite{Lin:2001zk,Lin:2004en,AMPTcode} 
has only quarks but no gluons, where the gluon degree of freedom can be considered as being absorbed in the quark's. Note that the scattering cross-sections in the parton cascade are set to be the same regardless of quark flavors.
Figure~\ref{fig:uds} shows the $v_2$ and $v_3$ of the $u$ and $d$ light quarks and the $s$ strange quarks in three systems: \AuAu\ and \dAu\ collisions at 200~GeV, and \pPb\ collisions at 5~TeV. The quark and antiquark anisotropies are found to be the same, so they are combined. There is practically no difference between the $u$ and $d$ quark $v_n$'s, so they are also combined in Fig.~\ref{fig:uds}. The $v_n$ magnitudes are similar among the three systems, except $v_3$ in \dAu\ which is significantly lower than the other two systems. In general small systems should generate lower $v_n$ than large systems, and this is the case for $v_3$ between \dAu\ and \AuAu\ collisions. The $v_2$ in \dAu\ is not much smaller than that in \AuAu, possibly because the lower energy density in \dAu\ is compensated by the larger elliptical eccentricity ($\epsilon_2$). The $v_n$ in \pPb\ are not much smaller than those in \AuAu, and this may be because the smaller system size is compensated by the larger collision energy.

At low $\pt$ the light quark $v_2$ is larger than the $s$ quark's. This is qualitatively consistent with the hydrodynamic picture where particles move with a common collective flow velocity. Because particles have the same $v_n$ at the same speed, $v_n(\pt)$ as a function of $\pt$ are split according to particle masses. This mass splitting between light and strange quarks is observed in both $v_2$ and $v_3$ and in all three systems. 

Since our previous studies~\cite{He:2015hfa,Lin:2015ucn} have shown that $v_2$ comes largely from the anisotropic escape mechanism, then the question is whether or not the observed mass splitting is entirely due to the minor contributions from hydrodynamics. So we also carry out a test calculation with no collective anisotropic flow by randomizing the outgoing parton azimuthal directions after each parton-parton scattering as in Ref.~\cite{He:2015hfa}. The results are shown by the dashed curves in Fig.~\ref{fig:uds}, where the differences between the light quark and strange quark $v_n$'s are still present. Since the parton azimuthal angles are now randomized, the final-state parton anisotropy is entirely due to the anisotropic escape mechanism~\cite{He:2015hfa}. 
The fact that the mass splitting is similar between the normal and \phirndm\ AMPT suggests that it is caused by the mass or kinematic difference in the scatterings rather than the collective flow. At high $\pt$ the light quark and strange quark $v_n$'s approach each other; this is expected because the mass difference becomes unimportant at high $\pt$.

\section{Mass splitting from quark coalescence}

\begin{figure*}[hbt]
  \begin{center}
    \includegraphics[width=\textwidth]{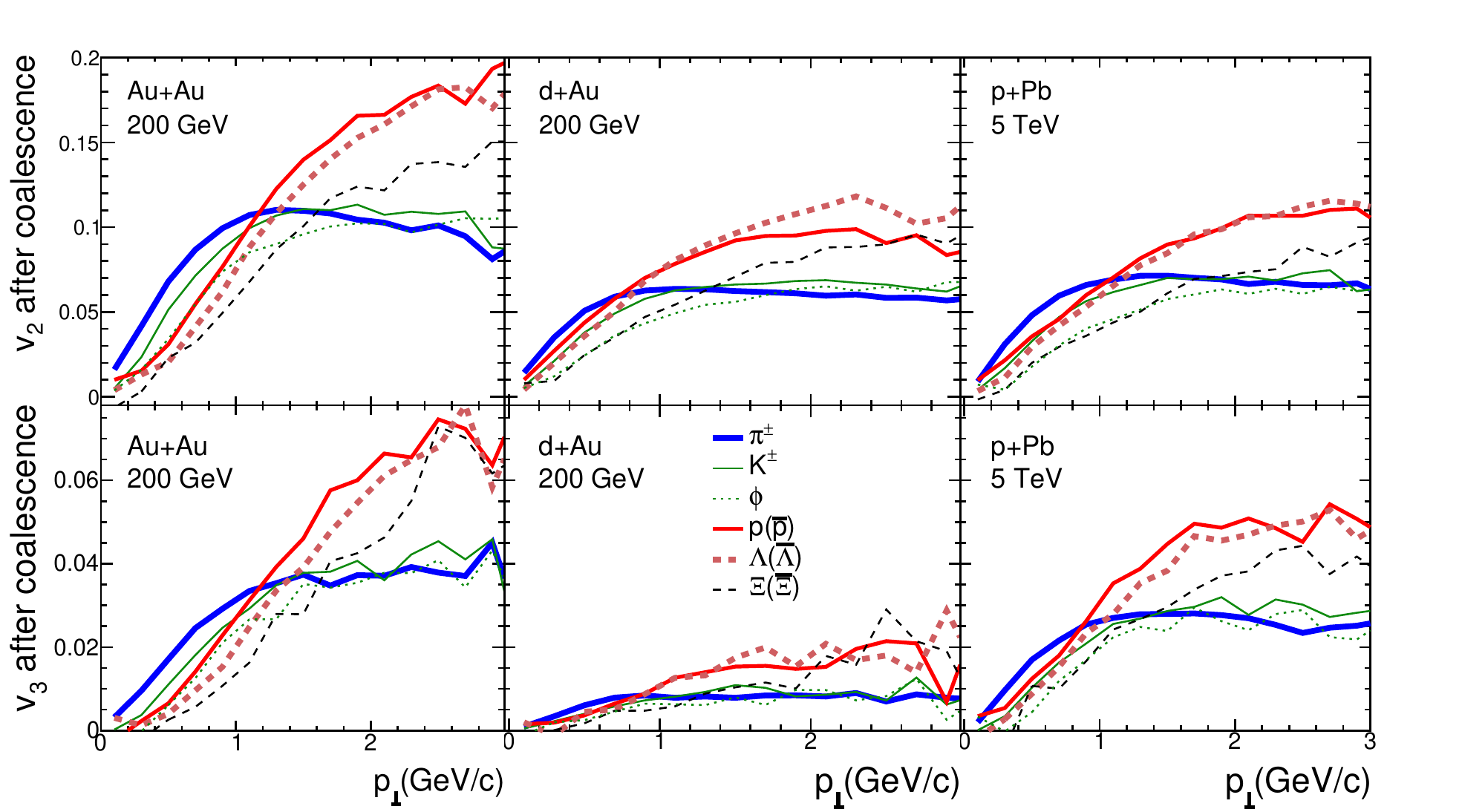}
    \caption{(Color online) {\em Mass splitting from coalescence.} Primordial hadron $v_2$ (upper panels) and $v_3$ (lower panels) as a function of $\pt$ right after quark coalescence but before hadronic rescatterings take place in AMPT with string melting. Three systems are shown: \bAu\ \AuAu\ collisions at $\snn=200$~GeV (left column), \bzero\ \dAu\ collisions at 200~GeV (middle column), and \bzero\ \pPb\ at 5~TeV (right column). Thick solid curves are for charged pions, thin solid curves for charged kaons, thin dashed curves for $\phi$-mesons, medium thick solid curves for (anti-)protons, thick dashed curves for $\Lambda$($\bar{\Lambda}$), and medium thick dashed curves for $\Xi$($\bar{\Xi}$).} 
    \label{fig:pikp}
  \end{center}
\end{figure*}

Since there is mass splitting in the quark $v_n$, it is natural to expect mass splitting in the $v_n$ of hadrons with different quark contents. 
However, for hadrons such as pions, $\rho$-mesons, and protons made of light quarks only, 
the difference between their anisotropies must come from the hadronization process and/or hadronic rescatterings. We first study the effect of the former by examining $v_2$ of hadrons right after hadronization but before hadron rescatterings take place. 
Figure~\ref{fig:pikp} shows the $v_2$ and $v_3$ of primordial $\pi$, $K$, $\phi$, $p$($\pbar$), $\Lambda$($\bar{\Lambda}$), $\Xi$($\bar{\Xi}$) as a function of $\pt$ in the three systems we studied. Note that primordial hadrons are hadrons formed directly from hadronization but before resonance decays and hadronic scatterings. 
In \AuAu\ collisions the particle $v_n$ exhibit the familiar mass-ordering at low $\pt$: the $v_n$'s of pions are larger than those of kaons which are in turn larger than those of (anti-)protons and strange baryons. The mass splittings in the small systems of \dAu\ and \pPb\ are not necessarily the same ordering as in the \AuAu\ system. In this section we study how this mass splitting comes about. We will concentrate on $v_2$ but the discussions can be extended to $v_3$.

Since the string melting version of AMPT forms hadrons via quark coalescence,  
the difference in the pion and proton $v_2$ comes from the difference in the number of constituent quarks and in the kinematics of those quarks. At high $\pt$ the hadron $v_2$ has been measured to exhibit the number of constituent quark (NCQ) scaling:
\begin{equation}
v_2^{\rm B}/3 \approx v_2^{\rm M}/2\;,
\end{equation}
where the superscripts `B' stands for baryons and `M' for mesons.
This comes naturally from quark coalescence, where two or three relatively high $\pt$ quarks are almost collimated and coalesce into a meson or baryon. The meson and baryon take on twice and three times the quark $v_2$ (which are saturated at high $\pt$ as in Fig.~\ref{fig:uds}), respectively. This NCQ scaling is evident in Fig.~\ref{fig:pikp}; the baryons in each graph approach a similar magnitude of $v_n$ in the higher $\pt$ region. 

\begin{figure}[hbt]
  \begin{center}
    \includegraphics[width=\columnwidth]{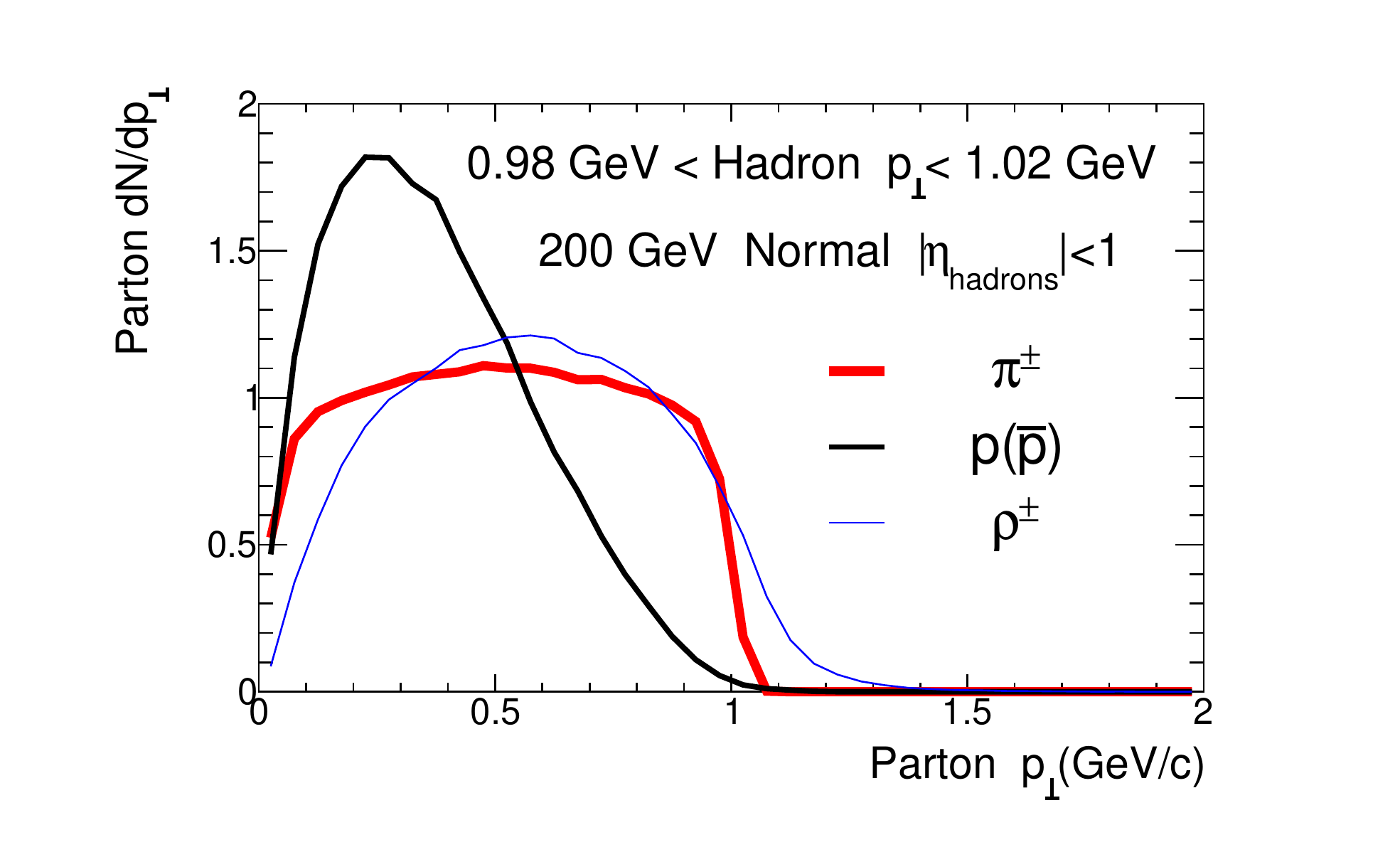}
    \includegraphics[width=\columnwidth]{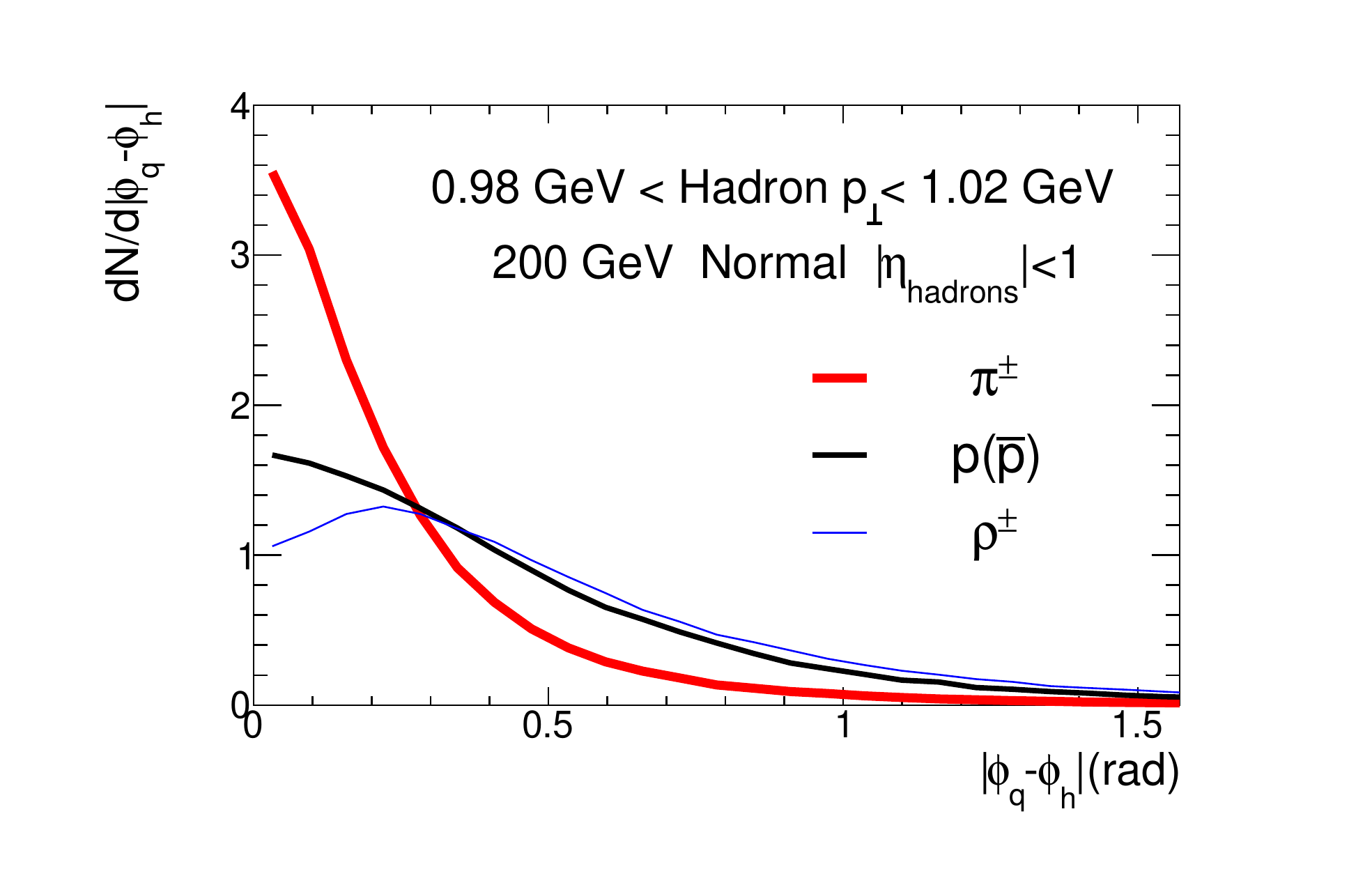}
    \caption{(Color online) {\em Coalescence kinematics.} Final-state $\pt$ (upper panel) and azimuthal opening angle (lower panel) distributions of constituent (anti-)quarks forming $\pi$, $\rho$, and (anti-)proton. Shown are string melting AMPT results for $b=$6.6-8.1~fm \AuAu\ collisions at 200~GeV.}
    \label{fig:kine}
  \end{center}
\end{figure}

However, this quark collimation picture cannot be extended to low $\pt$, since there the relative momentum among constituent quarks could be comparable to the hadron $\pt$, i.e. there will be finite opening angles among the constituent quarks. 
Therefore the kinematics in the coalescence process \cite{Lin:2009tk} such as finite opening angles could lead to the mass splitting of $v_2$ at low $\pt$.
To quantitatively understand this, we show in the upper panel of Fig.~\ref{fig:kine} the $\pt$ distributions for partons coalescing into pions and protons of $\pt=1$ \gevc. We have also depicted in the plots the $\rho$ meson, which has the same constituent quark content as the $\pi$ but a larger  mass. The lower panel of Fig.~\ref{fig:kine} shows the absolute difference between the azimuthal angle of the constituent quark and that of the formed hadron, $\dphi=|\phi_q-\phi_h|$.
Because of the finite angles, the average $\pt$ of the constituent quarks is larger than one half (one third) of the pion (proton) $\pt$. While the actual kinematics are complex, one may verify that a pair (or triplet) of partons with an average transverse momentum $\pt^{\rm q}$ at the average opening angle (as in Fig.~\ref{fig:kine}) gives the composite hadron $\pt^{\rm h}$ roughly as
\begin{equation}
  \pt^{\rm h}\approx\ncq\pt^{\rm q}\cos(\dphi)\;,
\end{equation}
where the superscripts `h' and `q' stand for hadrons and constituent quarks, respectively, and $\ncq$ is the number of constituent quarks for the given hadron type.

\begin{figure}[hbt]
  \begin{center}
    \includegraphics[width=\columnwidth]{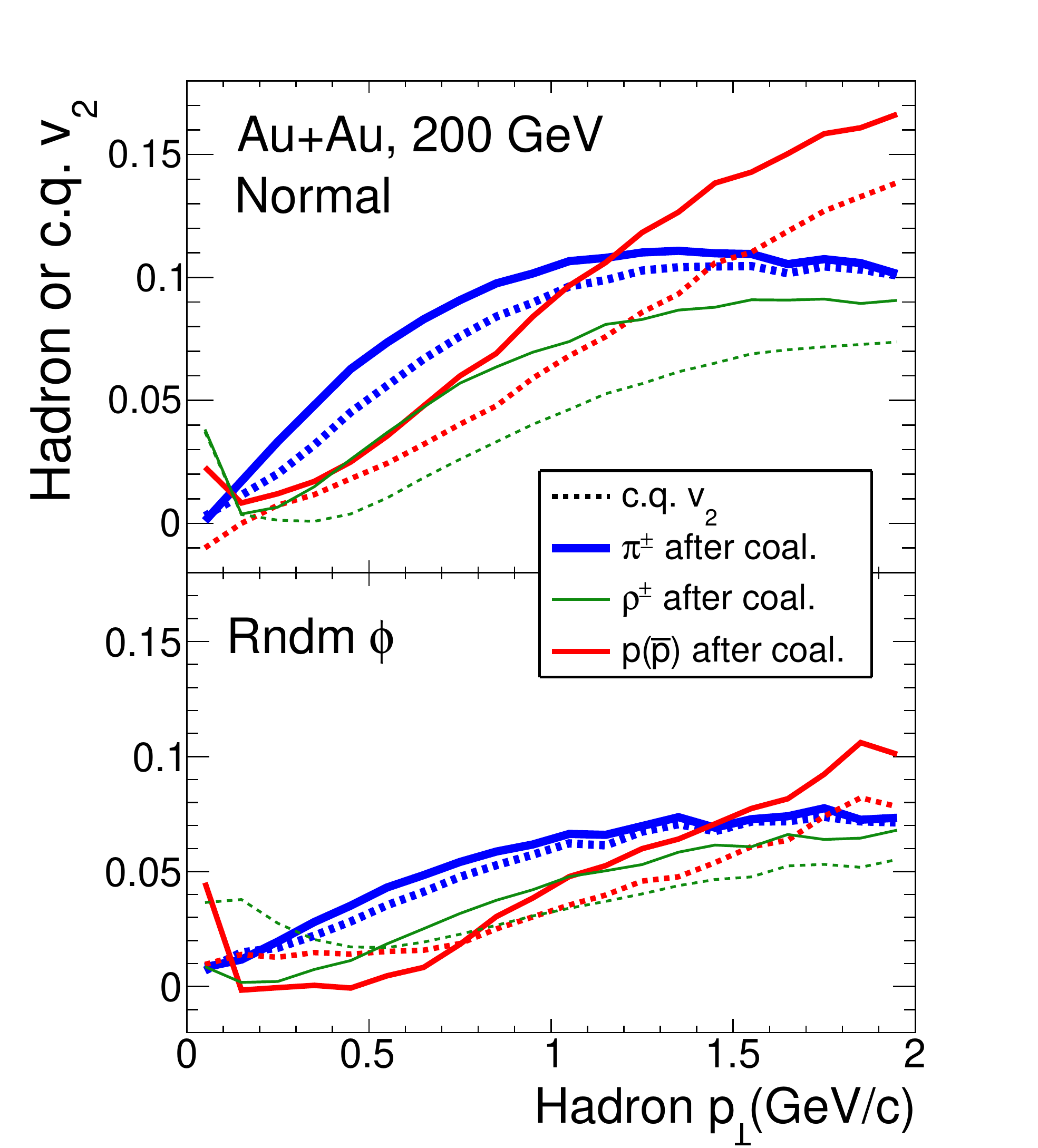}
    \caption{(Color online) {\em Conversion of constituent quark $v_2$ into hadron $v_2$ by coalescence.} Primordial hadron and constituent quark $v_2$, both plotted as a function of the {\em hadron} $\pt$. The hadron $v_2$ is taken before any hadronic rescatterings and the constituent quark $v_2$ is taken just before coalescence. Shown are both normal (upper panel) and \phirndm\ (lower panel) string melting AMPT results for \bAu\ \AuAu\ collisions at 200~GeV.}
    \label{fig:coal}
  \end{center}
\end{figure}

Similarly, because of the finite opening angle the hadron $v_2$ is not simply twice (or three times) the average quark $v_2$ at the corresponding average quark $\pt^{\rm q}$. This is shown in Fig.~\ref{fig:coal}, where the quark $v_2$ is plotted at the $\pt$ of the hadron it coalesces into, together with the hadron $v_2$ from Fig.~\ref{fig:pikp}. Note that the quark $v_2$ in Fig.~\ref{fig:uds} includes all quarks (i.e. from all hadrons) while that in Fig.~\ref{fig:coal} is categorized by the formed hadrons. As seen from Fig.~\ref{fig:coal}, the hadron $v_2$'s shown in solid curves are smaller than twice (three times) the quarks shown in dashed curves. Note that the shapes of the quark $v_2$ curves are different from each other because they are plotted at the hadron $\pt$ and because $\pt$ samplings of quarks into pions, $\rho$ mesons, and protons are different (c.f.~Fig.~\ref{fig:kine}). One may get a semiquantitative understanding of the hadron $v_2(\pt)$ curve by, again, using the average quark kinematics. The hadron azimuthal distribution is 
\begin{equation}
\Pi_{i=1}^{\ncq}[1+2v_2^{\rm q}\cos (2\phi_{{\rm q},i})]
\approx 1+2n_q v_2^{\rm q}\cos (2\dphi) \cos (2\phi_{\rm h})\;. 
\end{equation}
Thus the hadron $v_2$ is given by
\begin{equation}
v_2^{\rm h}(\pt^{\rm h})=\ncq v_2^{\rm q}(\pt^{\rm q})\cos (2\dphi)\;.\label{eq:vhvq}
\end{equation}
One may verify that this relationship, with the kinematics in Fig.~\ref{fig:kine}, can approximately describe the $v_2$ relationship between a hadron at $\pt=1$~\gevc\ and its constituent quarks in Fig.~\ref{fig:coal}.

Although $v_2$ is largely from the escape mechanism, there does exist a contribution from hydrodynamics in AMPT~\cite{He:2015hfa,Lin:2015ucn}. Thus we also carry out the test calculations with no collective anisotropic flow by randomizing the outgoing parton azimuthal directions after each parton-parton scattering as in Ref.~\cite{He:2015hfa,Lin:2015ucn}. The results are shown in the lower panel of Fig.~\ref{fig:coal}. In the \phirndm\ case, the final-state freezeout anisotropy is entirely due to the anisotropic escape mechanism. 
Since mass splitting is also observed in the randomized case, it indicates that the hydrodynamical collective flow is not required to generate the mass splitting in $v_n$ right after hadronization.

\section{Effects of resonance decays}

\begin{figure}[bht]
  \begin{center}
    \includegraphics[width=\columnwidth]{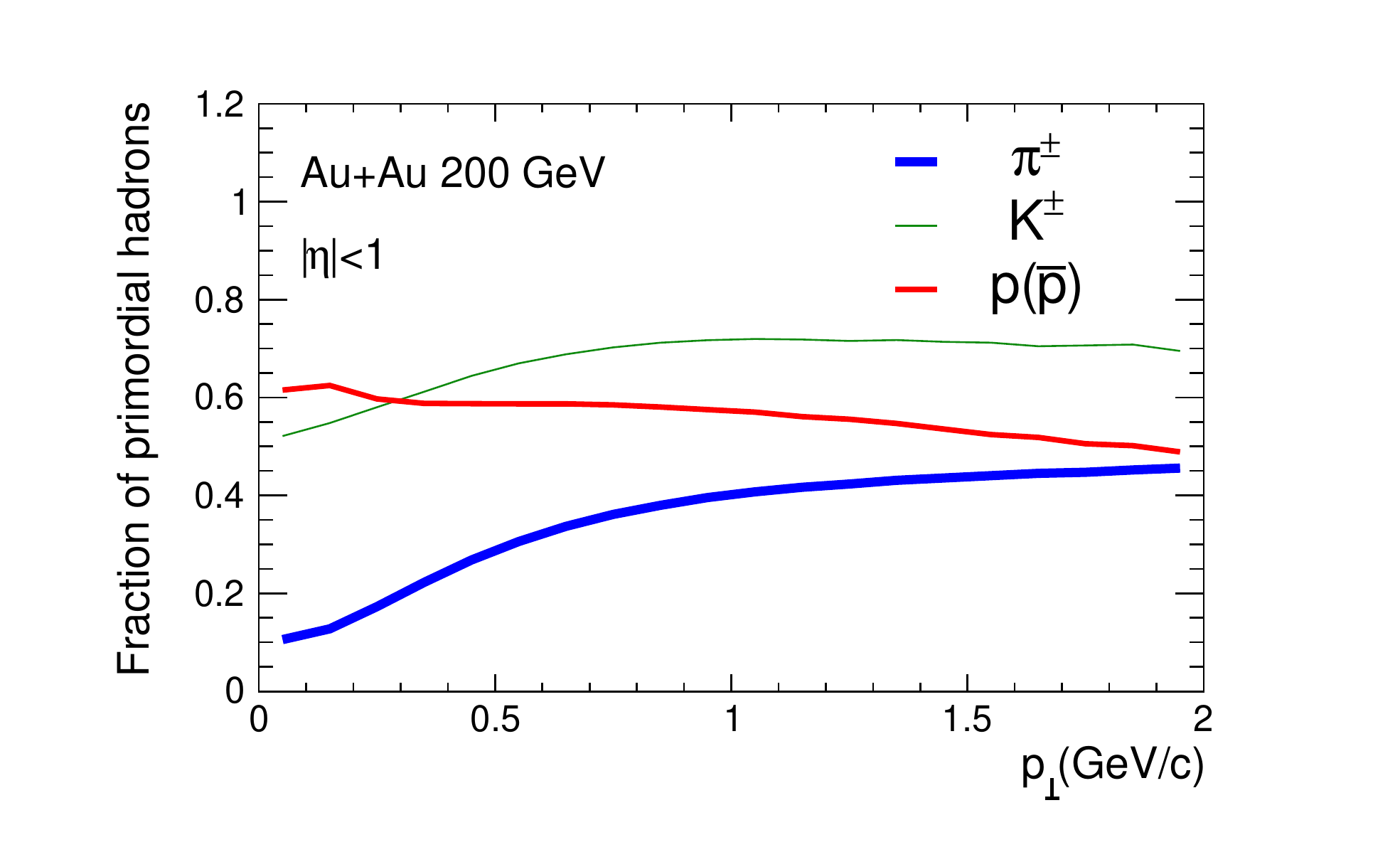}
    \caption{(Color online) {\em Decay contributions.} Fraction of primordial hadrons in \bAu\ \AuAu\ collisions at~200 GeV from string melting AMPT where hadronic rescatterings are turned off. 
}
    \label{fig:decay_frac}
  \end{center}
\end{figure}

What are shown in Fig.~\ref{fig:coal} are the $v_2$ values of primordial hadrons (obtained right after the quark coalescence in the AMPT evolution), not those of hadrons after resonance decays. 
Figure~\ref{fig:decay_frac} shows the fraction of primordial pions, kaons, and protons as a function of $\pt$. Since what we measure in detectors are particles after strong decays, we need to include the effects of resonances decays on $v_n$. 
In this section, we thus set the maximum hadronic stage to $\tmax=0.6$~\fmc\ in AMPT (parameter NTMAX=3) which turns off hadronic rescatterings. We then obtain the final-state hadron $v_n$ that include decays. 
Note that the final freezeout particles in AMPT include all strong decays of resonances but no electromagnetic or weak decays by default (except for the $\Sigma^0$ decay in order to include its feed down to $\Lambda$)~\cite{Lin:2004en}.
 
\begin{figure*}[hbt]
  \begin{center}
    \includegraphics[width=\textwidth]{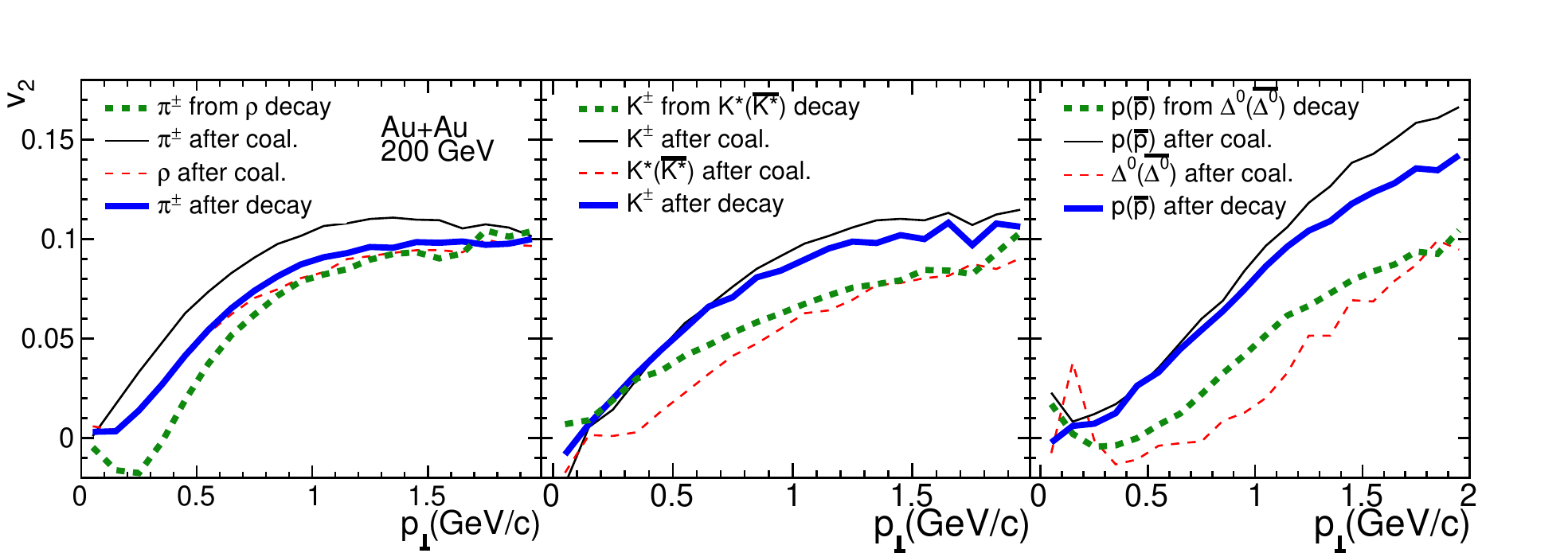}
    \caption{(Color online) {\em Effect of decays on $v_2$.} The $v_2$ of primordial hadrons (thin solid curves), resonances (thin dashed curves), decay products (thick dashed curves), and the total $v_2$ including both the primordial hadrons and decay products (thick solid curves). Shown are \bAu\ \AuAu\ collisions at~200 GeV from string melting AMPT where hadronic scatterings are turned off.}
    \label{fig:decay_effect}
  \end{center}
%\end{figure*}
%\begin{figure*}[hbt]
  \begin{center}
    \includegraphics[width=\textwidth]{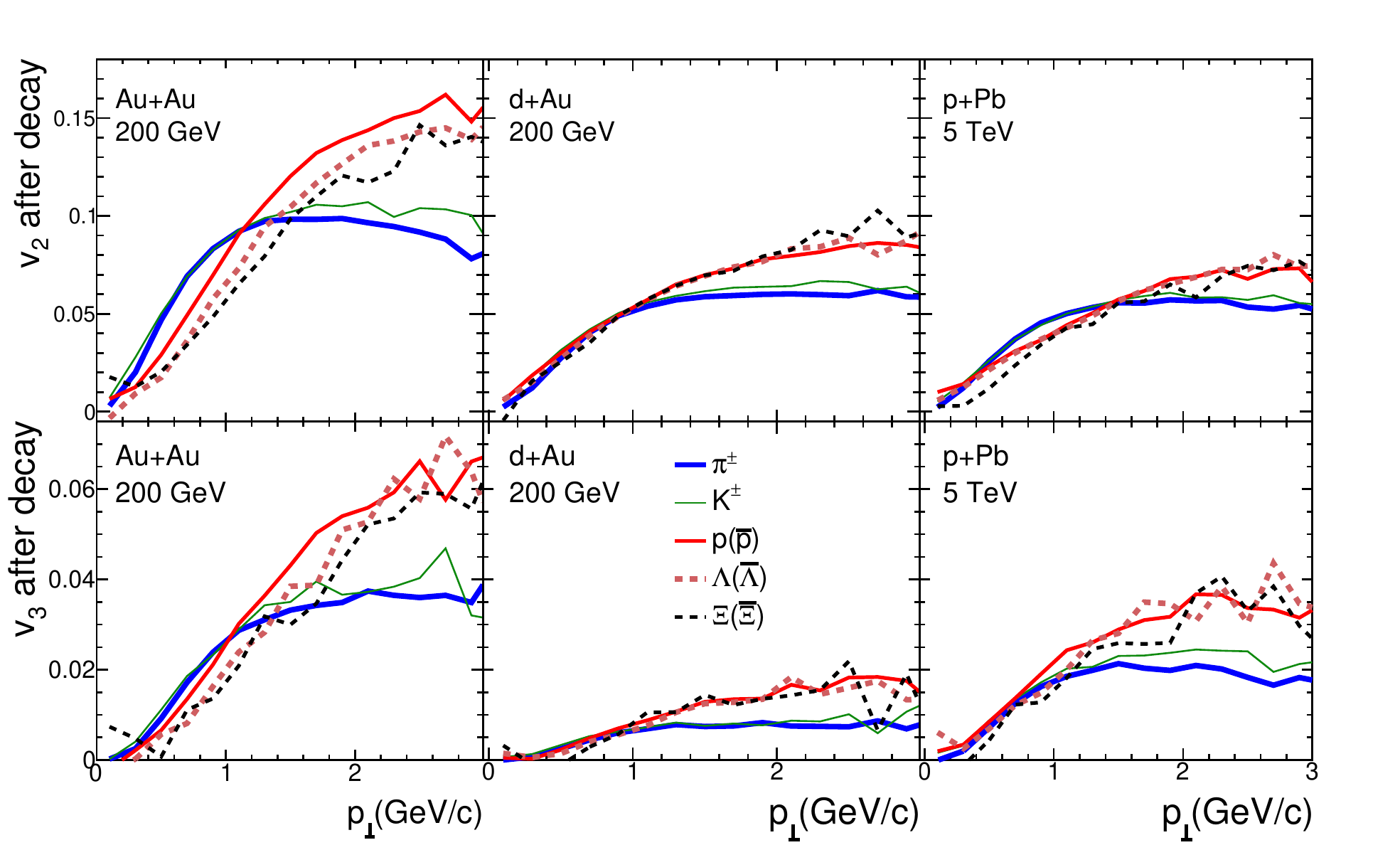}
    \caption{(Color online) {\em Effect of decays on mass splitting.} Hadron $v_2$ (upper panels) and $v_3$ (lower panels) including both primordial hadrons and decay products. Three systems are shown: \bAu\ \AuAu\ collisions at $\snn=200$~GeV (left column), \bzero\ \dAu\ collisions at 200~GeV (middle column), and \bzero\ \pPb\ at 5~TeV (right column). Results are from string melting AMPT where hadronic rescatterings are turned off. Thick solid curves are for charged pions, thin solid curves for charged kaons, medium thick solid curves for (anti-)protons, thick dashed curves for $\Lambda$($\bar{\Lambda}$), and medium thick dashed curves for $\Xi$($\bar{\Xi}$).}
    \label{fig:decay}
  \end{center}
\end{figure*}

The left panel of Fig.~\ref{fig:decay_effect} shows the $v_2$ of primordial pions, primordial $\rho$'s, pions from $\rho$ decays, and all pions. The middle panel shows the corresponding results for kaons where the $K^*$ decay channel is studied. The right panel shows 
the $v_2$ of primordial (anti-)protons, primordial (anti-)$\Delta^0$'s (as an example), protons from  (anti-)$\Delta^0$ decays (as an example), and all protons. 
We see that at low $\pt$ heavier primordial particles have smaller $v_2$. In addition, the decay product $v_2$ is usually smaller than their parent $v_2$. As a result, the $v_2$'s of final-state hadrons including the decay products are smaller than (or closely follow) those of the primordial particles. 
This reduction effect is stronger in pions than protons, because a bigger fraction of pions comes from resonance decays than protons according to Fig.~\ref{fig:decay_frac} and because the protons retain more of the parent $v_2$ than pions due to kinematics. 

Our results generally agree with those in Refs.~\cite{Eyyubova:2009hh,Qiu:2012tm,Crkovska:2016flo}.
%It is interesting to note that the $K^*$-decay kaon and $\Delta$-decay proton $v_2$ curves are shifted towards lower $\pt$ from their respective parent $v_2$ curve, while the $\rho$-decay pion $v_2$ is the opposite and the low-$\pt$ pions have negative $v_2$. This is because in the $K^*$ ($\Delta$) decay the daughter kaon (proton) follows mostly the parent direction, carrying a good fraction of the parent momentum and $v_2$, thus the daughter $v_2$ curve shifts towards left. On the other hand, the
It is interesting to note that the $\rho$-decay pion $v_2$ curve %relative to the parent $v_2$ is qualitatively different from those for the $K^*$ and $\Delta$ decays. This is because both decay pions are included in the former and only one daughter for the latter. 
at low $\pt$ does not follow the trend of the parent $\rho$-meson $v_2$.
Since the decay pion momentum in the $\rho$ rest frame is about 0.36~\gevc, in order to have a low-$\pt$ daughter pion in the lab frame, the $\rho$ decay must be very asymmetric: one pion at low $\pt$ and the other at high $\pt$. The high-$\pt$ pion closely follows the parent $\rho$ direction, while the low-$\pt$ pion aligns more perpendicularly due to momentum conservation. With positive $v_2$ of the $\rho$, there are therefore relatively more low-$\pt$ decay pions perpendicular to the reaction plane, hence a negative pion $v_2$. We have verified that this feature is also true for the pions from $\Delta$ decays.

Figure~\ref{fig:decay} shows the hadron $v_2$ and $v_3$ as a function of $\pt$ including contributions from resonance decays. The reduction in $v_n$ is evident in Fig.~\ref{fig:decay} in comparison with Fig.~\ref{fig:pikp} where only the primordial hadron $v_n$'s are shown. Because of the larger reduction in pion $v_n$ than in proton $v_n$ due to decays, the amount of mass-splitting is reduced. Depending on the magnitude of this reduction, the mass splitting between primordial hadrons right after coalescence may or may not survive once including the decay products. So in general the mass splitting effect decreases after including the decay products.

\section{Mass splitting from hadronic rescatterings}

Another source of mass splitting of $v_n$ comes from hadronic rescatterings. 
In the following we study $v_n$ as a function of the degree of hadronic rescatterings. We achieve this by varying the maximum allowed time, $\tmax$, of the hadronic interaction stage in AMPT. So $\tmax$ can be considered as a qualitative indicator of the amount of hadronic rescatterings. Note that there is no cut-off time for the partonic evolution in AMPT.

\begin{figure*}[hbt]
  \begin{center}
    \includegraphics[width=\textwidth]{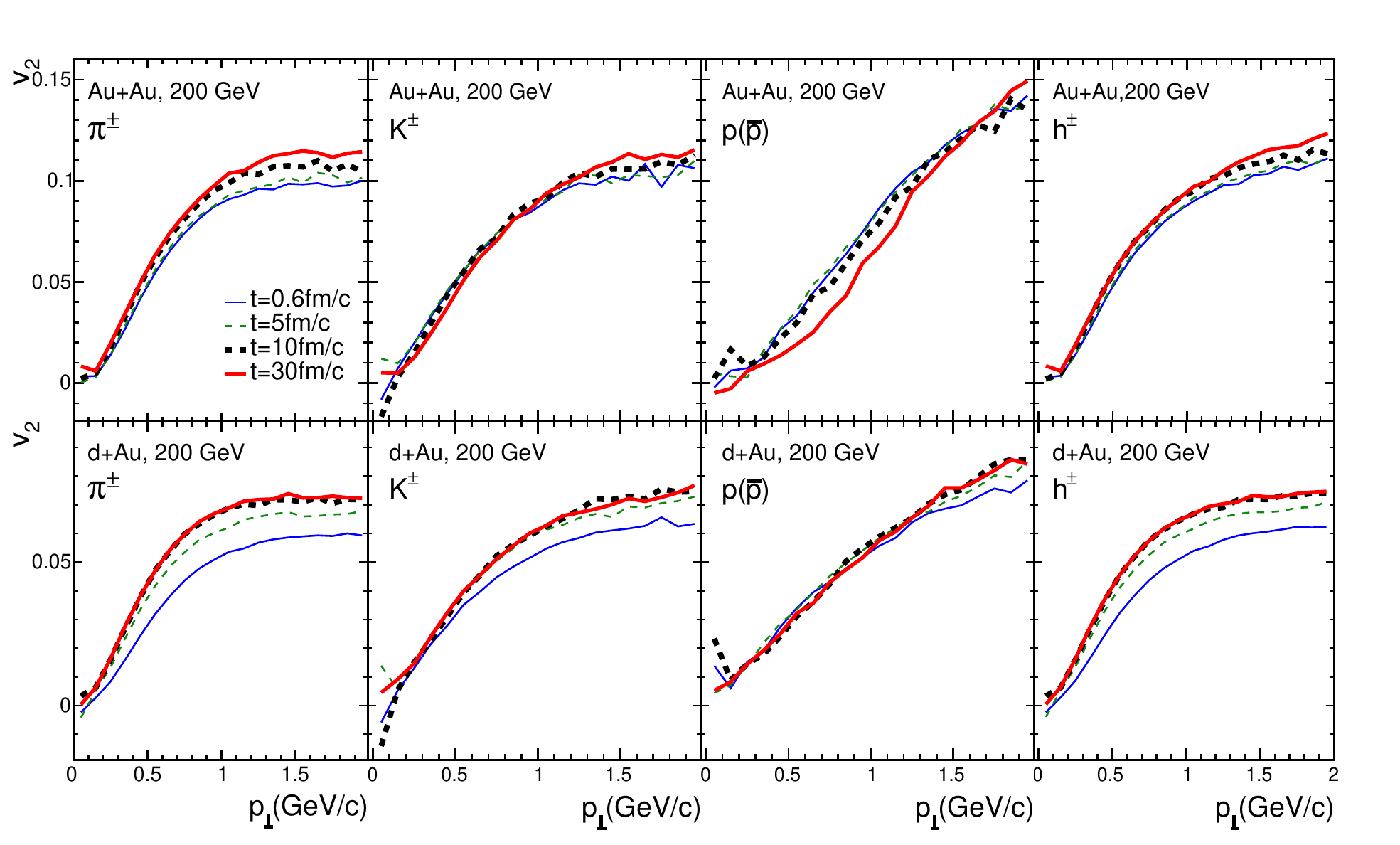}
    \caption{(Color online) {\em Effects of hadronic rescatterings on $v_2$.} Hadron $v_2$ as a function of $\pt$ at different stages of hadronic rescatterings in string melting AMPT. The $v_2$ are for final-state hadrons including resonance decays, where the final freezeout is controlled by the maximum allowed time ($\tmax$) for the hadronic stage. Shown are the results for charged pions (first column), charged kaons (second column), (anti-)protons (third column), and charged hadrons (last column) in \bAu\ \AuAu\ collisions at 200~GeV (upper panels) and \bzero\ \dAu\ collisions at 200~GeV (lower panels).}
    \label{fig:rescatt}
  \end{center}
\end{figure*}

The upper panels of Fig.~\ref{fig:rescatt} show the $v_2$ of charged pions, charged kaons, (anti)protons and charged hadrons (here defined as the sum of charged pions, kaons, protons and antiprotons) at freezeout in mid-central \AuAu\ collisions versus $\pt$ for various $\tmax$ values. We see that the pion $v_2$ increases with the amount of rescatterings, the proton $v_2$ decreases, while the kaon $v_2$ does not change significantly. 
This can be understood as the consequence of hadron interactions. 
For example, pions and protons tend to flow together at the same velocity
due to their interactions. Thus, pions and protons at the same velocity (i.e. small $\pt$ pions and large $\pt$ protons) will tend to have the same anisotropy after rescatterings.
This will then lead to lower $v_2$ for protons and higher $v_2$ for pions at the same $\pt$ value. 
Similar conclusions were reached in previous hadron cascade studies~\cite{Burau:2004ev,Petersen:2006vm,Zhou:2015iba} and a recent study with free-streaming evolution coupled to a hadron cascade~\cite{Romatschke:2015dha}.

Figure~\ref{fig:rescatt} also shows a small increase in the overall charged hadron $v_2$, and this is due to the remaining finite configuration space eccentricity before hadronic scatterings take place. 
In general, whether there is an overall gain in the $v_2$ of charged hadrons  depends on the configuration geometry at the beginning of hadron cascade. 
The lower panels of Fig.~\ref{fig:rescatt} shows our results for \dAu\ collisions.
We see that the pion $v_2$ increases significantly with hadronic scatterings while the proton $v_2$ remains roughly unchanged. 
Note that the overall gain in the charged hadron $v_2$ is larger in \dAu\ than \AuAu\ collisions, and this is due to the larger eccentricity in the \dAu\ system at the start of hadron cascade. Therefore the changes in the pion and proton $v_2$ are a net effect of the mass splitting due to pion-proton interactions (i.e. increase in the pion $v_2$ and decrease in the proton $v_2$) and the overall gain of $v_2$ for charged hadrons. 

As can be seen in Fig.~\ref{fig:rescatt}, $v_2$ continues to develop after hadronization in \AuAu\ as well as \dAu\ collisions. In \AuAu\ collisions the development happens mainly during 5-20~\fmc\, while in \dAu\ collisions the development happens earlier (mainly before 5~\fmc).
The spatial anisotropy is self-quenched due to the expansion and the development of momentum space anisotropy. 
The further increase of overall charged hadron $v_2$ in Fig.~\ref{fig:rescatt} suggests that the spatial anisotropy is not completely quenched at the time right after hadronization; a finite spatial anisotropy is present at the beginning of hadronic rescatterings which results in the further development of $v_n$. 

\begin{figure}[hbt]
  \begin{center}
    \includegraphics[width=\columnwidth]{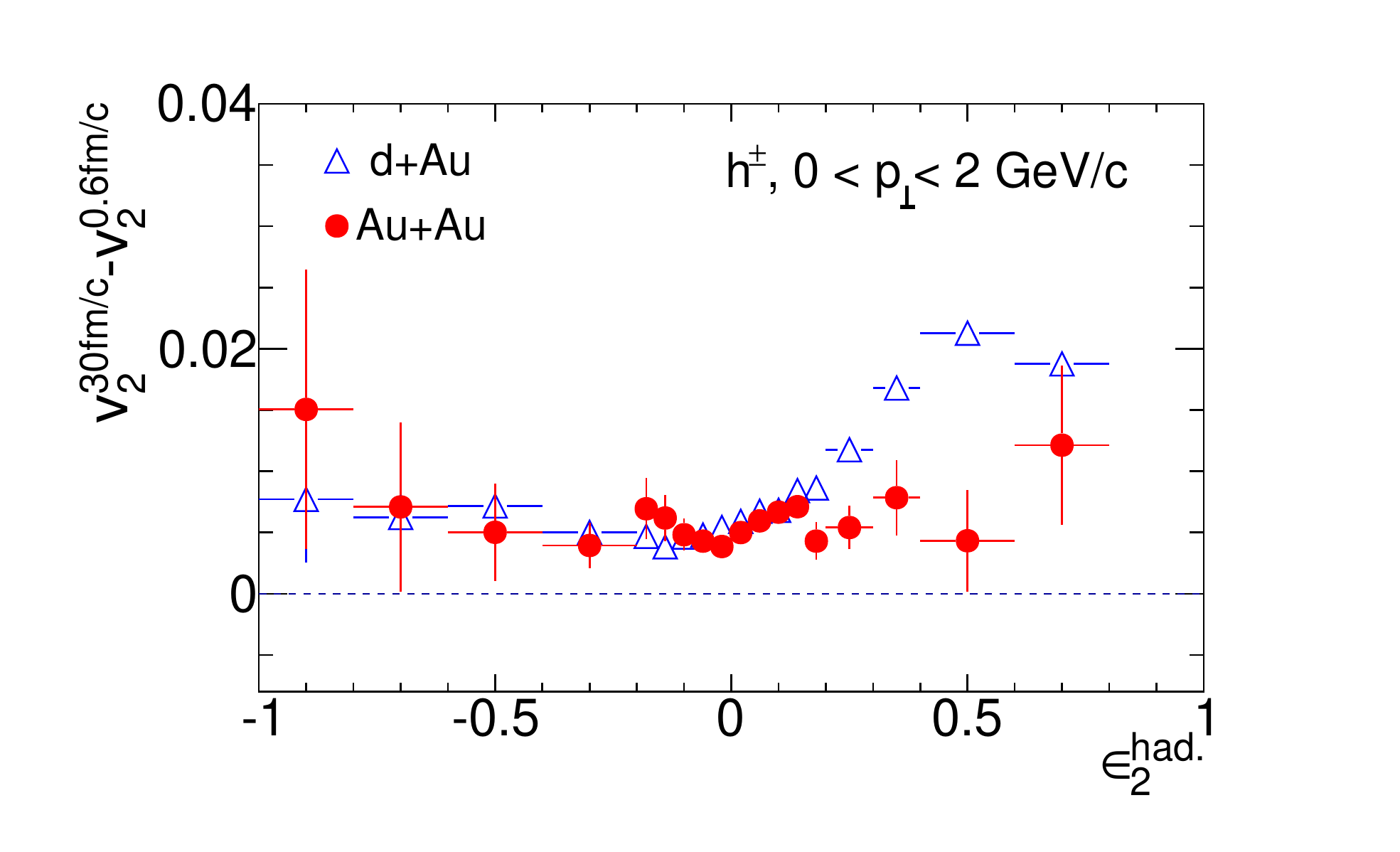}
    \caption{(Color online) {\em Connection to the initial hadronic eccentricity.} Gain in charged hadron $v_2$ due to hadronic rescatterings from hadronization ($\tmax=0.6$~\fmc) to final freezeout ($\tmax=30$~\fmc) as a function of the configuration space eccentricity of hadrons right after hadronization ($\ehad$) in \bAu\ \AuAu\ (circles) and \bzero\ \dAu\ (triangles) collisions at 200~GeV. The $v_2$ are for final-state hadrons including resonance decays.}
    \label{fig:dv2e2}
  \end{center}
\end{figure}

We elaborate this further by examining the $v_2$ increase as a function of the remaining eccentricity after hadronization ($\ehad$), i.e.~the starting eccentricity for hadronic cascade. 
This is shown in Fig.~\ref{fig:dv2e2} for both \AuAu\ and \dAu\ collisions. 
Since a typical AMPT evolution around mid-rapidity essentially ends by the time of 30~\fmc, we 
evaluate the increase in $v_2$ from hadronic scatterings as 
$\dv=\vfin-\vini$. The $\ehad$ value is calculated with respect to the initial configuration space $\psi_2^{(r)}$, as is $v_2$.
We have verified that the hadron $v_2$ right after the coalescence hadronization, as well as the $v_2$ at final freezeout, is proportional to the initial eccentricity ($\eini$)--which is also calculated with respect to the initial $\psi_2^{(r)}$--except when $\eini$ is large (close to one). 
We have also found that the $\ehad$ value is positively correlated with the $\eini$ value in \AuAu\ collisions, while the correlation is weak in \dAu\ collisions. 

Figure~\ref{fig:dv2e2} show that, in the $\ehad$ range of 0-0.2 in \AuAu\ and 0-0.5 in \dAu\ collisions, $\dv$ roughly increases linearly with $\ehad$. At large positive $\ehad$, the statistics are poor and $\ehad$ may not reflect a bulk geometry any more. At negative $\ehad$ events are also rare. 
On average, $\mean{\ehad}$ is 0.11 in \AuAu\ and 0.42 in \dAu\ collisions, starting from an initial $\mean{\eini}$ of 0.29 and 0.53 for the two collision systems, respectively. The geometric anisotropy is thus not quenched completely after partonic interactions in \AuAu\ collisions; the reduction in eccentricity in \dAu\ collisions is even smaller due to a shorter partonic stage. 
The remaining spatial anisotropy is smaller in \AuAu\ than in \dAu\ collisions, and this results in a smaller $v_2$ gain during the hadronic rescattering stage in \AuAu\ than in \dAu\ collisions, as observed in Fig.~\ref{fig:rescatt}.

It is also interesting to note in Fig.~\ref{fig:dv2e2} that $\dv$ is finite for events with $\ehad=0$, where one would naively expect $\dv=0$. This would indeed be true if the initial hadron $\vini$ (before hadronic rescatterings) was zero, analogous to the zero initial parton anisotropies $v_n^{\rm ini}\equiv0$ in AMPT (before partonic scatterings). However, for finite initial $\vini>0$, which is the case here, it is not necessarily true that $v_2$ would not further develop. 

\begin{figure*}[hbt]
  \begin{center}
    \includegraphics[width=\textwidth]{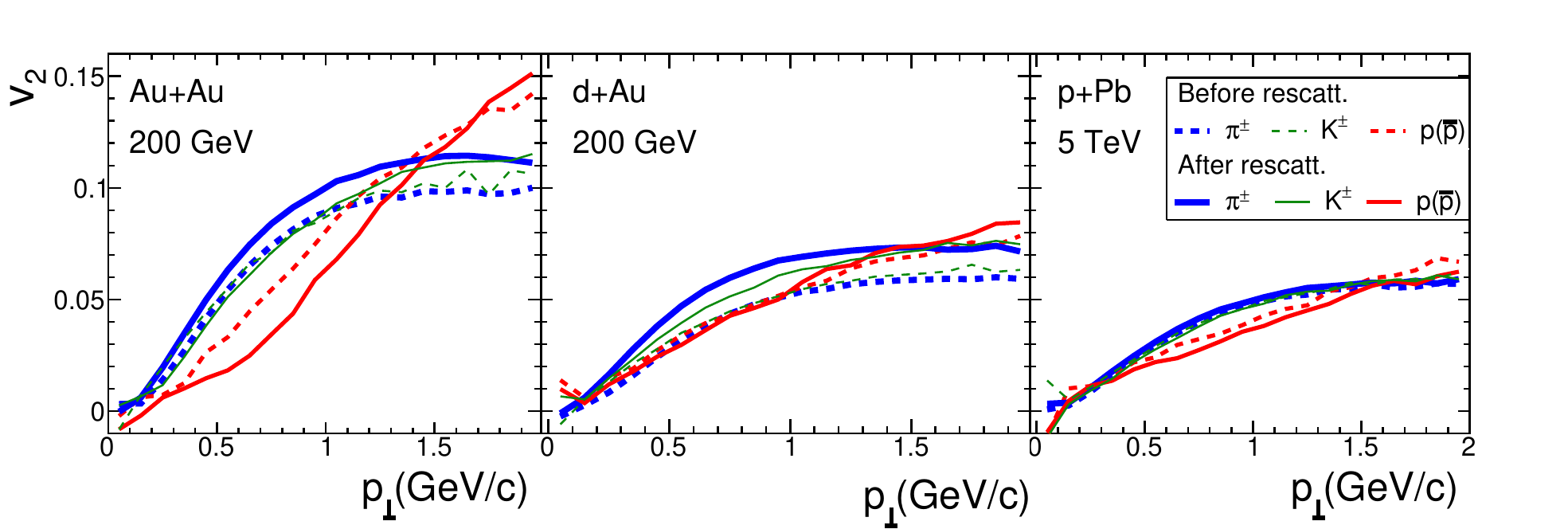}
    \caption{(Color online) {\em Effect of hadronic rescatterings on mass splitting.} 
      Charged pions (thick curves), charged kaons (thin curves), and (anti-)proton (medium thin curves) $v_2$ as a function of $\pt$ before (dashed curves) and after (solid curves) hadron rescatterings in string melting AMPT. The effects of resonance decays are included. Three systems are shown: \bAu\ \AuAu\ collisions at $\snn=200$~GeV (left column), \bzero\ \dAu\ collisions at 200~GeV (middle column), and \bzero\ \pPb\ at 5~TeV (right column).}
    \label{fig:rescatt_split}
  \end{center}
\end{figure*}

Figure~\ref{fig:rescatt_split} shows hadron $v_2$ as a function of $\pt$ before hadronic rescatterings but including resonance decays in dashed curves and $v_2$ of freezeout hadrons after hadronic rescatterings in solid curves. As shown, hadronic rescatterings make significant contributions to the mass splitting in the final-state hadron $v_2$. Meanwhile the absolute gain of the $v_2$ magnitude is relatively small during the hadronic stage. 

\begin{figure*}[hbt]
  \begin{center}
    \includegraphics[width=\textwidth]{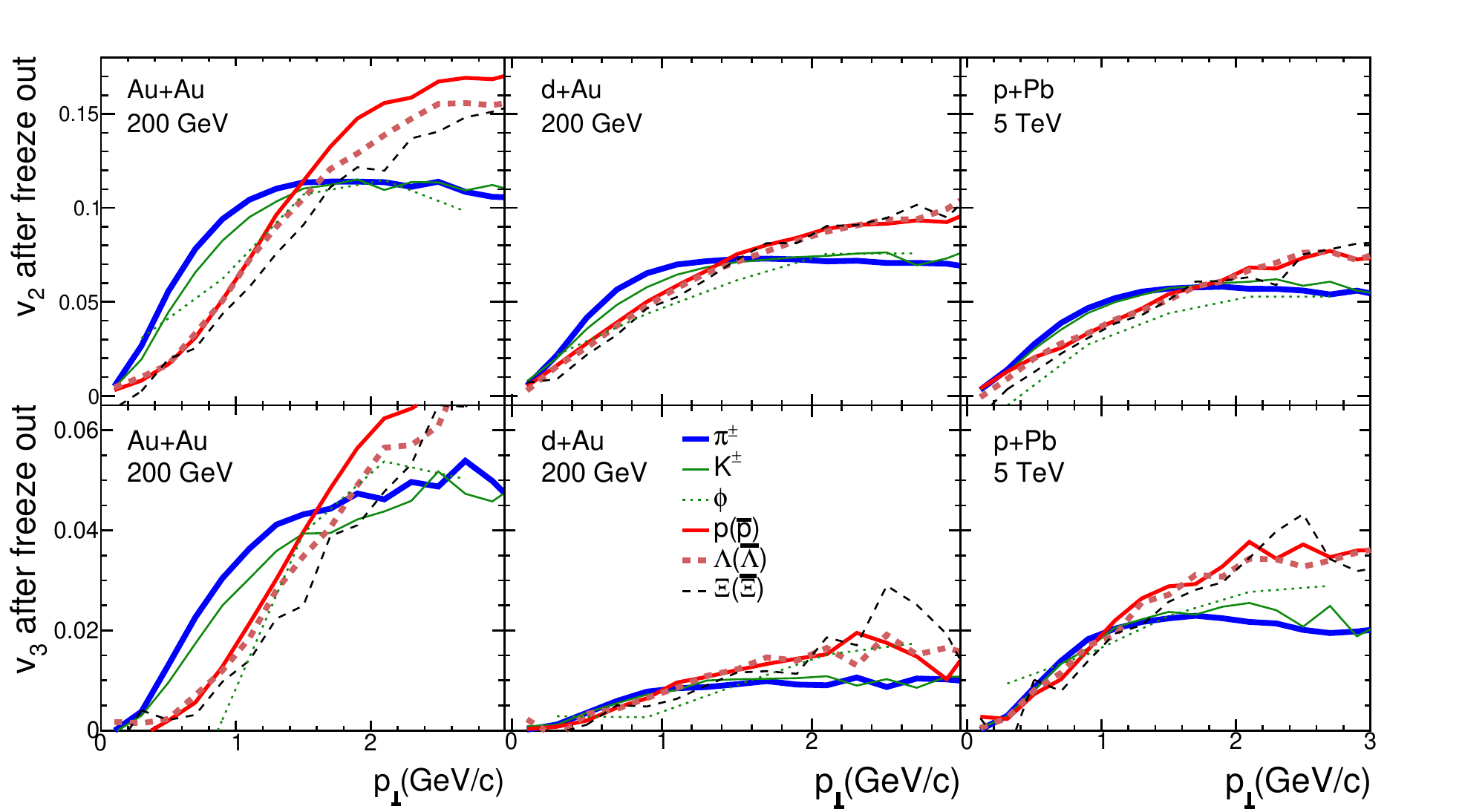}
    \caption{(Color online) {\em Mass splitting at freezeout.} Final hadron $v_2$ (upper panels) and $v_3$ (lower panels) as a function of $\pt$ from string melting AMPT, where resonance decays are included. Three systems are shown: \bAu\ \AuAu\ collisions at $\snn=200$~GeV (left column), \bzero\ \dAu\ collisions at 200~GeV (middle column), and \bzero\ \pPb\ at 5~TeV (right column). Thick solid curves are for charged pions, thin solid curves for charged kaons, thin dashed curves for $\phi$-mesons, medium thick solid curves for (anti-)protons, thick dashed curves for $\Lambda$($\bar{\Lambda}$), and medium thick dashed curves for $\Xi$($\bar{\Xi}$).}
    \label{fig:freezeout}
  \end{center}
\end{figure*}

%\section{Summary on the origins of mass splitting}
\section{Discussions}

Figure~\ref{fig:freezeout} shows $v_2$ and $v_3$ of final-state hadrons as a function of $\pt$ in \AuAu\ and \dAu\ collisions at 200~GeV and \pPb\ collisions at 5~TeV. The $\phi$-mesons are all decayed in the final state of AMPT, so they are reconstructed by the invariant mass of $K^+K^-$ pairs~\cite{Pal:2002aw} and $K_SK_L$ pairs with combinatorial background subtraction, as usually done in experiments~\cite{Abelev:2008aa}. The mass splitting of $v_n$ at low $\pt$ is more obvious in \AuAu\ collisions than small systems.
There is also splitting in $v_n$ at high $\pt$, likely more due to the number of constituent quarks 
rather than the mass difference. 

\begin{figure*}[hbt]
  \begin{center}
    \includegraphics[width=\textwidth]{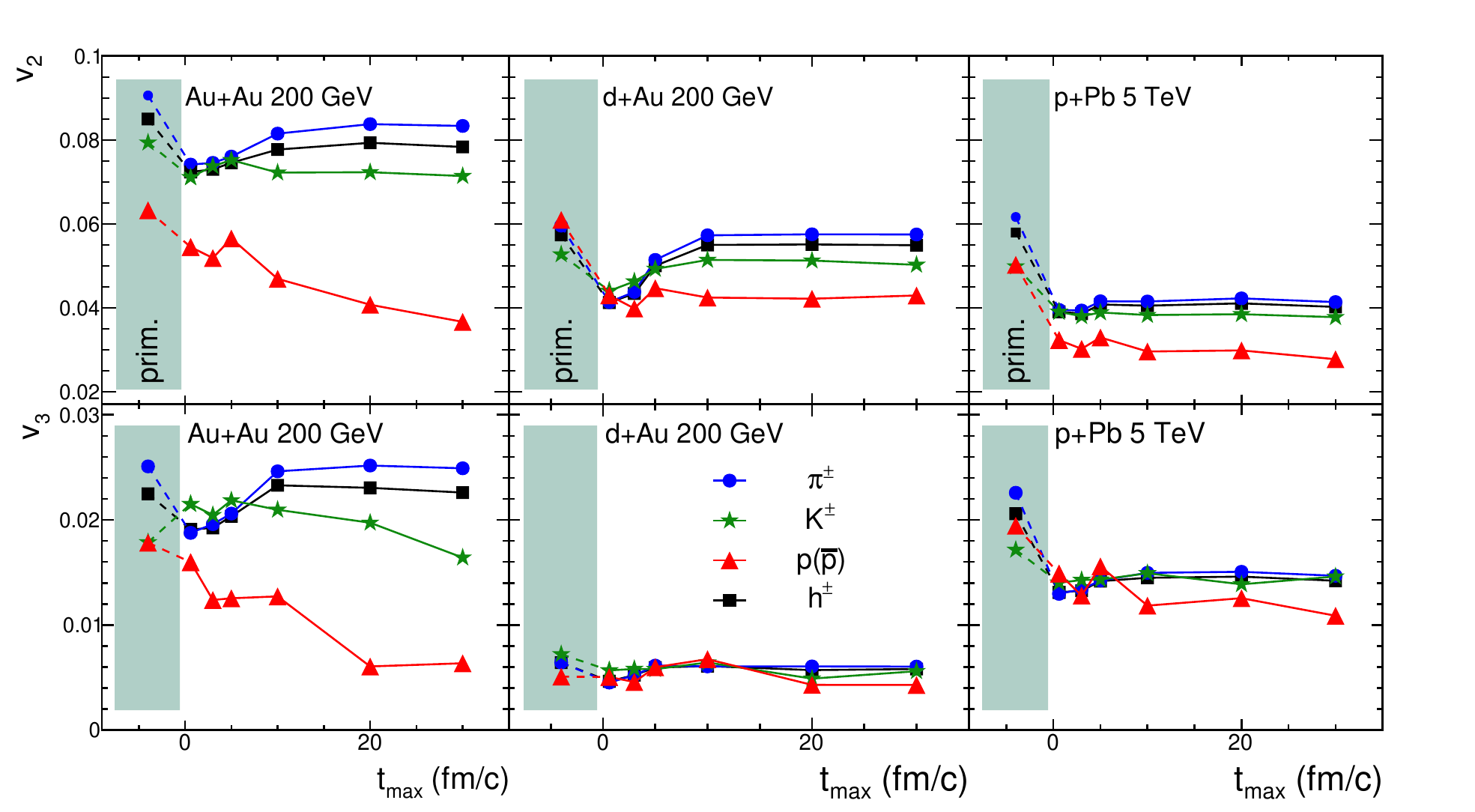}
    \caption{(Color online) {\em Origin of $v_n$ mass splitting.} The $v_2$ (upper panels) and $v_3$ (lower panels) of charged pions, charged kaons, (anti-)protons, and charged hadrons within $0.8<\pt<1.2$~\gevc\ at different stages of the collision evolution. Points plotted within the shaded areas represent primordial hadrons right after quark coalescence, points plotted at $\tmax=0.6$~\fmc\ represent results right after hadronization but including resonance decays, and points at various $\tmax$ values represent results after hadronic rescatterings (and resonance decays). }
    \label{fig:summary}
  \end{center}
\end{figure*}

We summarize our results on the mass splitting of $v_2$ with Fig.~\ref{fig:summary}, which shows the  $v_2$ of pions, kaons, protons and anti-protons, and charged hadrons 
within a fixed $\pt$ bin of $0.8<\pt<1.2$~\gevc, as an example. 
Different stages of the collision system evolution are shown: (i) right after the quark coalescence hadronization including only primordial particles (data points plotted to the left of $\tmax=0$); (ii) right after the quark coalescence but including resonance decays (data points plotted at $\tmax=0.6$~\fmc); (iii) after various degrees of hadronic rescatterings, which are obtained from freezeout particles by setting $\tmax$ to the corresponding values as plotted. As shown in Fig.~\ref{fig:summary}, most of the overall $v_n$ is built up in the partonic phase, while the additional gain in the overall $v_n$ from hadronic rescatterings is small. 
On the other hand, although there is often a significant mass splitting in the primordial hadron $v_n$ right after hadronization due to the kinematics in the quark coalescence process, the mass splitting is often reduced when decay products are included in $v_n$. In other words, the mass splitting before hadronic rescatterings is usually small. This small mass splitting does not change significantly during the first 5~\fmc\ in \AuAu\ collisions, since the partonic stage dominates the early evolution. We also see that a significant mass splitting is built up during the time of 5-20~\fmc\ of hadronic rescatterings. After 20~\fmc\ there is little further change in the $v_n$ in \dAu\ or \pPb\ collisions, while in \AuAu\ there is still a small increase in the size of mass splitting.

%%%%%%%%%%%%%%%%%%%%%%%%%%%%%%%%%%%%%%%%%%%%%%%%%%%%%%%%%%%
\section{Conclusions}

We have studied the developments of the mass splitting of hadron $v_n$ at different stages of nuclear collisions with a multi-phase transport model AMPT. 
First results on Au+Au and \dAu\ collisions at the top RHIC energy have been published in Ref.~\cite{Li:2016flp}. The present work provides extensive details to that earlier study by including more hadron species such as resonances and strange hadrons. We also expand the investigation to the triangular flow $v_3$ and \pPb\ collisions at the LHC energy of 5~TeV.
We reach the same conclusion for $v_2$ and $v_3$, for both heavy ion collisions and small-system collisions, in that the mass splitting of hadron $v_n$ is partly due to the quark coalescence hadronization process but more importantly due to hadronic rescatterings. 
Although the overall $v_n$ amplitude is dominantly developed during the partonic stage, the mass splitting is usually small right after hadronization, especially after including resonance decays. 
The majority of the hadron mass splitting is developed in the hadronic rescattering stage, even though the gain in the overall $v_n$ of charged particles is small there. 
These qualitative conclusions are the same as those from hybrid models that couple hydrodynamics to a hadron cascade, even though in transport models such as AMPT the anisotropic parton escape is the major source of $v_n$. 
In the \phirndm\ test of AMPT, where the anisotropic parton escape is the only source of $v_n$, we also observe similar mass splitting of hadron $v_n$.
Therefore we conclude that the mass splitting of $v_n$ can be an interplay of several physics processes and is not a unique signature of hydrodynamic collective flow.

%%%%%%%%%%%%%%%%%%%%%%%%%%%%%%%%%%%%%%%%%%%%%%%%%%%%%%%%%%%%%%%%%%%%%%
\section*{Acknowledgments}

This work is supported in part by US~Department of Energy Grant No.~DE-SC0012910 (LH,FW,WX), No.~DE-FG02-13ER16413 (DM), and the National Natural Science Foundation of China Grant No.~11628508 (ZWL) and No.~11647306 (FW). HL acknowledges financial support from the China Scholarship Council. 

\bibliographystyle{unsrt}
\bibliography{ref}
\end{document}